\documentclass[12pt,fleqn]{article}
\usepackage{latexsym}

\usepackage{graphicx}

\begin{document}
\title{How to Measure the Nonlocal Phase of a Single Photon}
\author{Noam Erez$\footnote{E-mail:nerez@physics.tamu.edu}$
 {\ }\\
\small  {\em \small Institute for Quantum Studies and
Department of Physics, Texas A\&M University,}\\ \small {College
Station, \small TX 77843-4242, USA}
}

\maketitle
\begin{abstract} The relative phase between spatially separated
 component waves of a single photon can be measured by joint interference
with a second photon emitted by a known source. In the case of a
single such phase (i.e. two component waves), the probability for a
successful measurement is one half. This method can be implemented
with current experimental techniques.


\end{abstract}

\vspace{.1 in}

\section{Introduction}

\begin{figure}
  \centering
  \includegraphics[width=0.5 \textwidth]{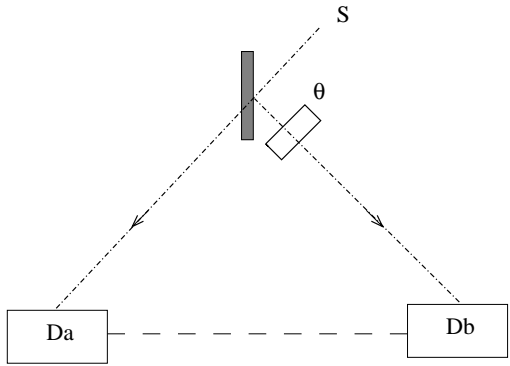}\\
  \caption{Measurement of nonlocal phase. Detectors $D_a,~D_b$ act locally, but may share entanglement
(Indicated by dashed line).}\label{fig1}
\end{figure}

Consider an ensemble of photons each separately impinging on the
beamsplitter of Fig.\ref{fig1}. Their state after leaving the
beamsplitter and the phaseshifter on the right, is
$|L\rangle+e^{i\theta}|R\rangle$, in an obvious notation. The
simplest way of measuring the relative phase, $\theta$, is to
recombine the beams in some common region, and let them interfere.
The phase is then determined in the usual way from the interference
pattern. This is essentially the principle of telescopy. If the
phase to be measured can take any value, a large number of
measurements will be needed to determine it with precision. On the
other hand, if we restrict the phase to either of two values
differing by $\pi$, the two states are mutually orthogonal
and an optimal measurement should be able to tell them apart every
time, obviating the need to consider an ensemble. We do not restrict
$\theta$ for now, for the sake of comparison with familiar single
photon interference.

Suppose, however, that we limit ourselves to measurements performed
locally on the two beams without recombining them. If the measuring
devices share no quantum correlations (entanglement), we can gain no
information on the relative phase. It is known, however, that if the
two observers share an EPR pair this can be done, in principle,
e.g., by placing an atom in each of the two remote locations, and
having the photon absorbed by the one it comes in contact with. This
is then followed by a Bell measurement on the two atoms\footnote{
After interacting with the atoms through the unitary transformation
corresponding to absorption with probability one:
\begin{equation}
\left(|k_1\rangle+e^{i\phi}|k_2\rangle\right)|g\rangle_L |g\rangle_R
\mapsto |0\rangle \left(|e_L,g_R\rangle+e^{i\phi}|g_L,e_R\rangle\right)
\end{equation}
The two atoms are now in a superposition of two Bell states.
Performing Bell measurements on the ensemble provides the phase. It
has been shown\cite{NL1}, that Bell measurements on such a system
can, in principle, be performed using local interactions and
entanglement.}.

As we shall see in the next section, the same goal can be accomplished much more simply,
albeit with half the efficiency, by producing a second such photon,
with known relative phase and recombining the two locally at each
side as shown in Fig.\ref{fig3}.

\section{Two photon interference}

\begin{figure}
  \centering
  \includegraphics[width=0.25\textwidth]{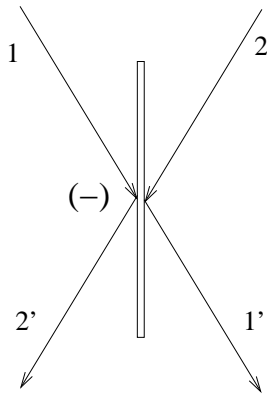}\\
  \caption{Beam splitter.}\label{fig2}
\end{figure}

Fig.\ref{fig2} depicts a beam splitter and two plane wave modes\footnote{The output of a beam
splitter of finite size cannot be described arbitrarily well by two plane waves (even when the input is such). 
At the end of this section
we shall briefly outline the treatment of a more realistic state of the form $|L\rangle+e^{i\theta}|
R\rangle$}. The beam splitter is assumed to be lossless and with coefficients of
reflection and transmission of equal magnitude (`a 50-50 beam
splitter'). The modes are symmetric with respect to reflection.
There is still some freedom in the choice of the phases of these
coefficients, and for definiteness we shall choose them to be real:

\begin{equation}
\left(
  \begin{array}{r}
    |1_{\textbf{k}_1} \rangle \\
    |1_{\textbf{k}_2} \rangle \\
  \end{array}
\right) \mapsto \frac{1}{\sqrt{2}}\left(
  \begin{array}{rr}
    1 & 1 \\
    -1 & 1 \\
  \end{array}
\right) \left(
  \begin{array}{r}
    |1_{\textbf{k}'_1} \rangle \\
    |1_{\textbf{k}'_2} \rangle \\
   \end{array} \right)
\end{equation}The interaction picture is implicitly assumed, so the only explicit
time evolution is that caused by the beam splitters.
Where we commit the common abuse of notation of denoting `output modes',
here distinguished by primes, where we really mean the same `input
modes' at a later time. This single photon scattering matrix is the
same as the classical one. It also gives the full quantum scattering
matrix in the Heisenberg representation\cite{QBS}:

\begin{equation}
\left(
  \begin{array}{r}
    a^\dag_{\textbf{k}_1} \\
    a^\dag_{\textbf{k}_2} \\
  \end{array}
\right) \mapsto \frac{1}{\sqrt{2}}\left(
  \begin{array}{rr}
    1 & 1 \\
    -1 & 1 \\
  \end{array}
\right) \left(
  \begin{array}{r}
    a^\dag_{\textbf{k}'_1} \\
    a^\dag_{\textbf{k}'_2} \\
  \end{array} \right) \label{QBSeq}
\end{equation}

The price we pay for the choice of real phases is an asymmetry, here
in the reflection coefficients, between the left and right modes.
This will be indicated in the figures by a minus sign on the side
where the reflection is accompanied by a $\pi$ phase shift.

\begin{figure}
   \centering
  \includegraphics[width=0.5 \textwidth]{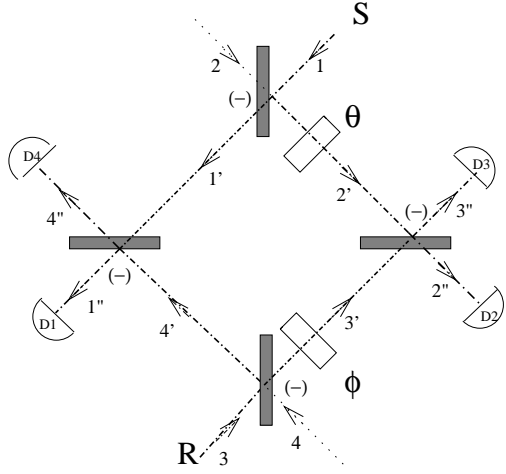}\\
  \caption{Set-up for detecting a photon's nonlocal phase by 2 photon interference.}\label{fig3}
\end{figure}

Fig.\ref{fig3} depicts the interference pattern of the ``source"
photon of Fig.\ref{fig1} in mode 1, with a ``reference" photon in mode
3. In other words, the input state is $|1_1,1_3\rangle$, where we
have dropped the $\mathbf{k}$s. The resulting interference pattern can be
calculated by applying Eq.\ref{QBSeq} once for each of the beam
splitters, in the appropriate order to $|\mathrm{in}\rangle
=a_1^\dag a_3^\dag |0\rangle$:

\begin{eqnarray}
 a_1^\dag a_3^\dag &\mapsto& \frac{1}{2}\left(a_{1'}^\dag+e^{i\theta}a_{2'}^\dag\right)
 \left(e^{i\phi}a_{3'}^\dag+a_{4'}^\dag\right) \nonumber \\
   &\mapsto& \left(\frac{1}{2}\right)^2\left[\left(a_{1''}^\dag +a_{4''}\right) +
   e^{i\theta}\left(a_{2''}^\dag -a_{3''}\right)\right]
   \left[e^{i\phi}\left(a_{2''}^\dag +a_{3''}\right) + \left(-a_{1''}^\dag +a_{4''}\right)\right] \nonumber \\
   &=& \frac{1}{4}\left[\left(a_{4''}^{\dag2}-a_{1''}^{\dag2}\right)-e^{i(\theta+\phi)}
   \left(a_{3''}^{\dag2}-a_{2''}^{\dag2}\right)\right] \nonumber \\
   &+& \frac{e^{i\theta}+e^{i\phi}}{4}\left(a_{1''}^\dag a_{3''}^\dag+a_{2''}^\dag a_{4''}^\dag
   \right)+\frac{e^{i\theta}-e^{i\phi}}{4}\left(a_{1''}^\dag a_{2''}^\dag+a_{3''}^\dag
   a_{4''}^\dag\right).
\end{eqnarray}Or,

\begin{eqnarray}
  |1_1,1_3\rangle &\mapsto& \frac{1}{\sqrt{8}}\left[|2_{4''}\rangle-|2_{1''}\rangle+
  e^{i(\theta+\phi)}\left(|2_{3''}\rangle-|2_{2''}\rangle \right)\right] \nonumber \\
   &+&\frac{e^{i\theta}+e^{i\phi}}{4}\left(|1_{1''},1_{3''}\rangle+|1_{2''},1_{4''}\rangle\right)
    +\frac{e^{i\theta}-e^{i\phi}}{4}\left(|1_{1''},1_{2''}\rangle+|1_{3''},1_{4''}\rangle\right).
\end{eqnarray}

Assuming perfectly efficient detectors, the probability of finding 2
photons at the same detector is\footnote{If the detectors do not
distinguish between one and two photons, but are 100\% efficient,
this corresponds to having only one detector click.} 1/2. This is
the famous Hong-Ou-Mandel ``bunching"\cite{HOM}. These events give
us no information on $\theta$. The probability of getting clicks in
either detectors 1 and 3, or 2 and 4; is $P(|1_1,1_3\rangle) =
\frac{1}{4}\cos^2\left(
\frac{\theta-\phi}{2}\right)=P(|1_2,1_4\rangle)$, and likewise
$P(|1_1,1_2\rangle) = P(|1_3,1_4\rangle) =\frac{1}{4}\sin^2\left(
\frac{\theta-\phi}{2}\right)$. The relative frequency of the
occurrence of $\{\{1,2\},\{3,4\}\}$ vs. $\{\{1,3\},\{2,4\}\}$
evidently gives us information on $\theta$. In particular, if
$\theta$ can take on two values:
$\theta\in\{\theta_0,\theta_0+\pi\}$, we can choose $\phi=\theta_0$
and get perfect distinguishability when the 2 photons exit different
beam splitters (i.e. half of the time).

\begin{figure}
  \centering
  \includegraphics[width=1.0 \textwidth]{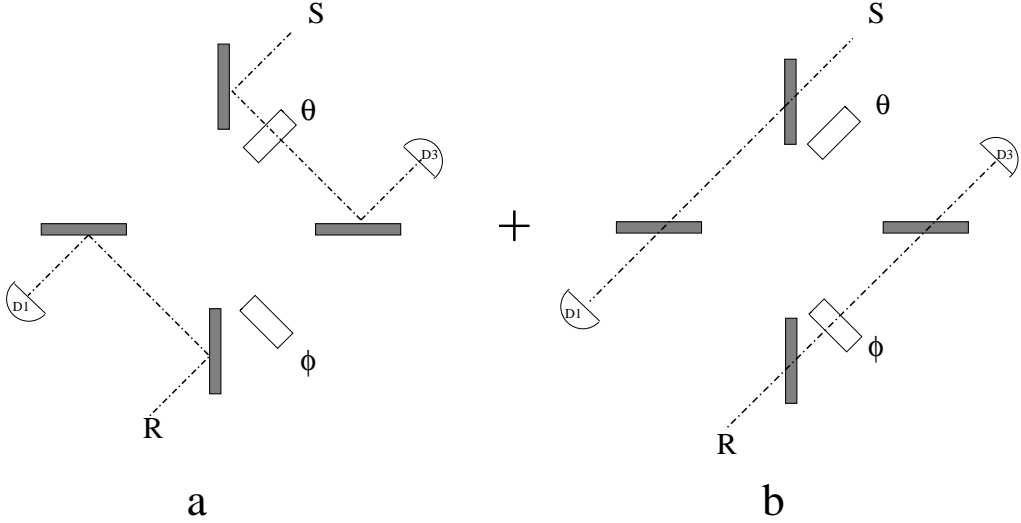}\\
  \caption{The two terms contributing to the output $|1_{1''},1_{3''}\rangle$.}\label{fig4}
\end{figure}

Fig.\ref{fig4} illustrates graphically the two two-photon contributions
to the probability amplitude, and hence to the probability of the
outcome $|1_{1''},1_{3''}\rangle$:

\begin{equation}
P\left(|1_{1''},1_{3''}\rangle\right)=\left|e^{i\theta}\left(\frac{1}{\sqrt{2}}\right)^4(-1)^2+
e^{i\phi}\left(\frac{1}{\sqrt{2}}\right)^4\right|^2
=\frac{1}{4}\cos^2\left(\frac{\theta-\phi}{2}\right).\end{equation}
The rest of the terms can be interpreted similarly.

\begin{figure}
  \centering
  \includegraphics[width=1.0 \textwidth]{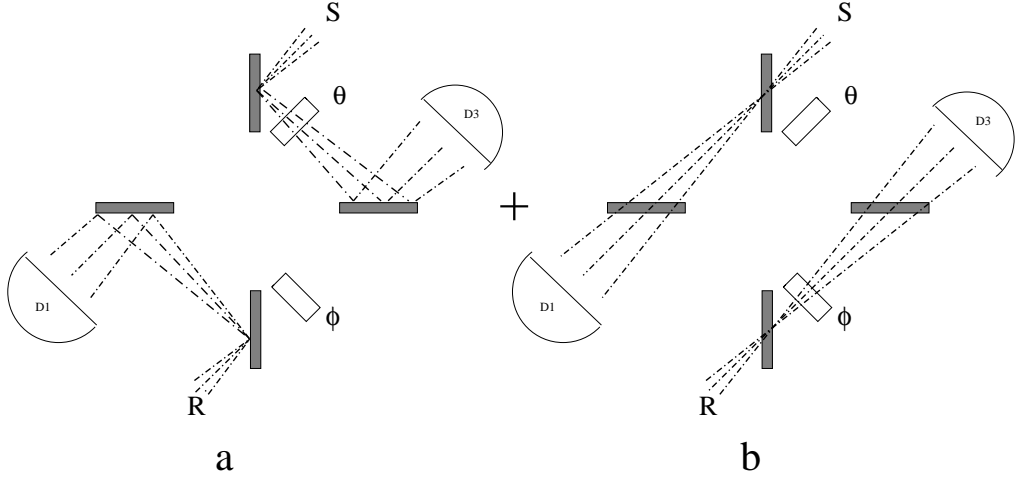}\\
  \caption{Two photon interference with non-planar waves. (Compare Fig.\ref{fig4})}\label{fig6}
\end{figure}

We are now ready to drop the assumption that $|L\rangle,~|R\rangle$ are plane
waves. As depicted in Fig.\ref{fig6}, we consider the case where the
primary and reference source each emit a superposition of several
plane waves. It is still assumed that the two sources emit a beam
with the same state, up to translation and rotation. The initial
state is now $|\mathrm{in}\rangle=\sum_i \alpha_i a^\dag_{1i} \sum_j \beta_j
a^\dag_{3j}|0\rangle$. The figure shows graphically the geometrical
optical argument demonstrating that the two pathways corresponding to the
photons reaching a given pair of detectors are indeed
indistinguishable and interfere.

\section{Multiple relative phases.}

\begin{figure}
  \centering
  \includegraphics[width=0.5 \textwidth]{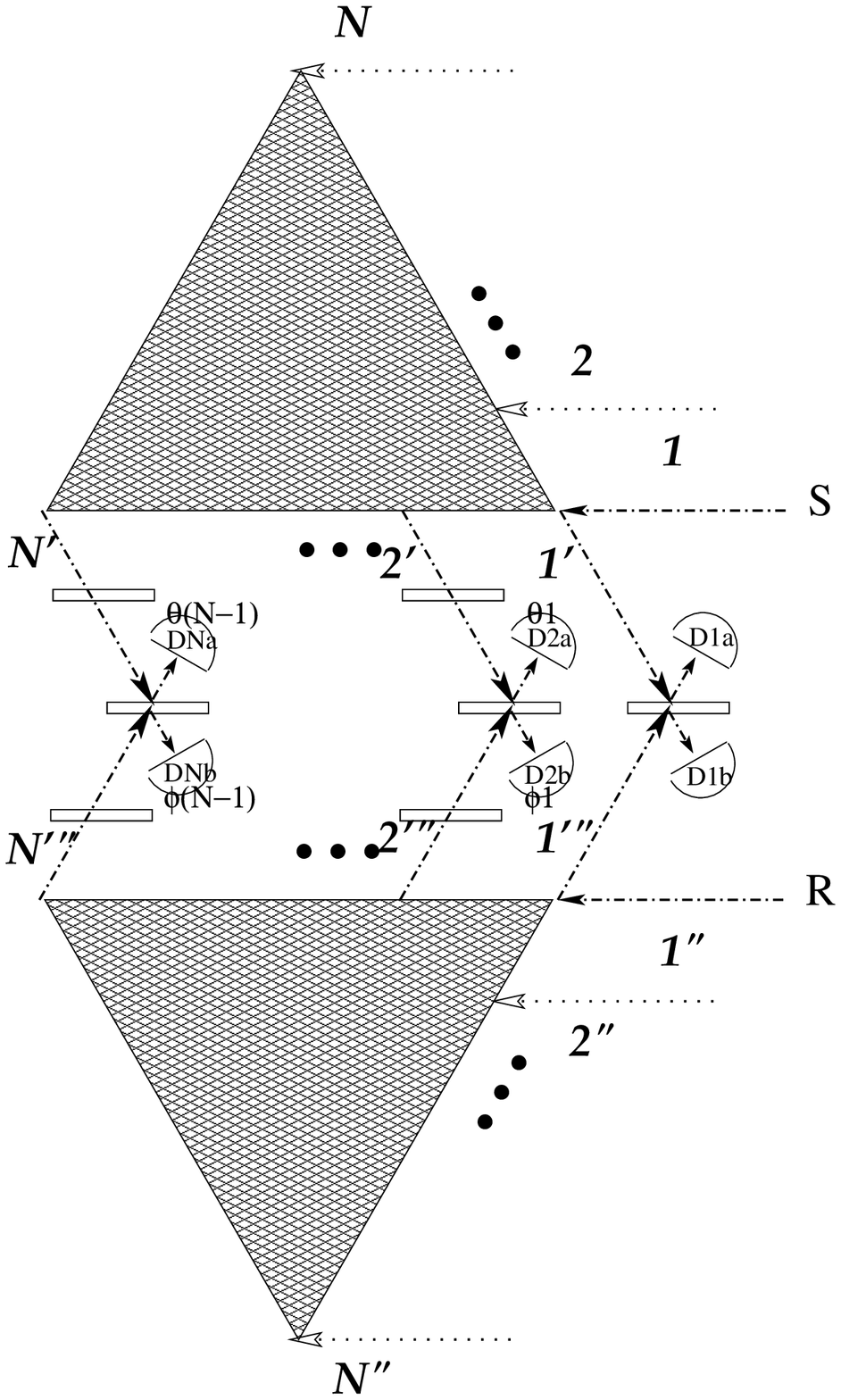}\\
  \caption{Two photon interference with N-1 independent relative phases. The large triangles denote
symmetric N-mode multiport beam
splitters\cite{tritters}.}\label{fig5}
\end{figure}

The fact that we recover, on average, only half a bit of information
on the local phase per photon, may seem troubling. Therefore, let us
consider the generalization to more than one non-local phase, in
which the situation is in some sense better.

Consider the states $|1_1\rangle+e^{i\theta_1}|1_2\rangle\ldots
+e^{i\theta_{N-1}}|1_N\rangle$ with $\theta_i\in
\{\theta_{i0},\theta_{i0}+\pi\}$. The relative phase between each
two terms again takes one of two values differing by $\pi$. If we
now apply the analysis of the previous section to the set-up
depicted in Fig. \ref{fig5}, and choose $\phi_i=\theta_{i0}$, we
will find that with probability $N^{-1}$ both photons will be
absorbed by the same detector, giving away no information on any of
the relative phases. With the complementary probability, we will get
clicks at two detectors placed next to two different beam splitters,
and the identity of these detectors will determine the relative
phase of the two appropriate terms of the source state. Thus, we
gain, on average, $1-N^{-1}$ bits, where 'average' means expectancy.
If we keep the same relative phases for different experimental runs,
then as the number of runs becomes comparable to $N$, we shall get a
lot of redundant information and the efficiency will go down.
Relaxing the condition that the relative phases take on pairs of
complimentary values is another possibility. We shall avoid these
combinatorial complications by simply assuming that the phases are
changed between experimental runs.

\section{Discussion}

As we have seen, by means of two photon interference, one can
measure the nonlocal phase, or phases, between spatially separated
components of a single photon. For $N$ components, one can gain on
average $1-N^{-1}$ bits of information on the nonlocal phases per
measurement. Recently, nonlocal measurements on a single photon
using homodyne detection were discussed \cite{homodyne}. It seems to
me that the simple scheme outlined above also sheds some light on
the latter.

I have been careful not to use the controversial term `the photon's
wavefunction' in the hope of avoiding a long digression. A good
discussion of this concept is found in \cite{SZ}.

As mentioned in the introduction, it is known that to measure a nonlocal property
of a system by local devices, they must share entanglement. It is
tempting in our present scheme to think of the photon of the
reference source as the part of the measuring device carrying the
entanglement. One might justify this distinction between the two
photons by arguing that their orthogonal states make them
distinguishable. From a quantum field point of view, a single
nonlocal photon is indeed an entangled state (between of the field
at various locations). As long as the two photons don't overlap, we
can think of the overall state as a product of the two single-photon
states. The only places where the two photons overlap, are precisely
in the vicinity of the local detectors.

Finally, note the similarity to Hanbury-Brown--Twiss\
effect\cite{HBT}. Here, however, one of the interfering photons
comes from a reference source.

\begin{center}
ACKNOWLEDGMENTS
\end{center}

It is a pleasure to acknowledge helpful conversations with Benni
Reznik. I gratefully acknowledge the support of the Welch founation
(grant no. A-1218).

\end{document}